\documentclass{article}

\usepackage[numbers]{natbib}
\usepackage{PRIMEarxiv}

\usepackage[utf8]{inputenc} % allow utf-8 input
\usepackage[T1]{fontenc}    % use 8-bit T1 fonts
\usepackage{hyperref}       % hyperlinks
\usepackage{url}            % simple URL typesetting
\usepackage{booktabs}       % professional-quality tables
\usepackage{amsfonts}       % blackboard math symbols
\usepackage{nicefrac}       % compact symbols for 1/2, etc.
\usepackage{microtype}      % microtypography
\usepackage{lipsum}
\usepackage{fancyhdr}       % header
\usepackage{graphicx}       % graphics
\graphicspath{{media/}}     % organize your images and other figures under media/ folder

\title{Machine learning operations: A survey on MLOps tool support
\thanks{\textit{\underline{Citation}}: 
\textbf{N. Hewage, D. Meedeniya, Machine Learning Operations: A Survey on MLOps Tool Support, 2022, pp. 1-12, Arxiv.2202.10169.  DOI:  	
https://doi.org/10.48550/arXiv.2202.10169 [online: http://arxiv.org/abs/2202.10169]}} 
}

\author{
  Nipuni Hewage \\
  Axiata Digital Labs\\
  Parkland Level 11, 33 Park Street\\ 
  Colombo 2\\
  Sri Lanka\\
  \texttt{nipunitharu35@gmail.com} \\
   \And
  Dulani Meedeniya\\
  Department of Computer Science and Engineering\\
  University of Moratuwa\\ Moratuwa\\ Sri Lanka\\
  \texttt{dulanim@cse.mrt.ac.lk} \\
}

\begin{document}
\maketitle

\begin{abstract}
Machine Learning (ML) has become a fast-growing, trending approach in solution development in practice. Deep Learning (DL) which is a subset of ML, learns using deep neural networks to simulate the human brain. It trains machines to learn techniques and processes individually using computer algorithms, which is also considered to be a role of Artificial Intelligence (AI). In this paper, we study current technical issues related to software development and delivery in organizations that work on ML projects. Therefore, the importance of the Machine Learning Operations (MLOps) concept, which can deliver appropriate solutions for such concerns, is discussed. We investigate commercially available MLOps tool support in software development. The comparison between MLOps tools analyzes the performance of each system and its use cases. Moreover, we examine the features and usability of MLOps tools to identify the most appropriate tool support for given scenarios. Finally, we recognize that there is a shortage in the availability of a fully functional MLOps platform on which processes can be automated by reducing human intervention.
\end{abstract}

% keywords can be removed

\section{Introduction}

At present, many tasks are related to software products or services to a certain extent. With the availability of a vast amount of data, most of the software solutions require functionalities such as data acquisition, processing, analysis, predictions and visualizations. Therefore, the software development process has been integrated with the data science technique to provide the required solutions. Different phases of the Software Development Life cycle (SDLC) produce a range of unique artefacts. With the increase of development iterations of a project, often there is a need for managing a larger number of artefacts generated through those iterations. Thus, more emphasis has given rapid changes and frequent integration to build the product incrementally. This approach is often referred to as Continuous Integration (CI) and Continuous Deployment (CD)\cite{34RubasingheICCIP2017}.

Consequently, CI along with the process-specific improvements such as Development and Operations (DevOps) practices have become one of the increasingly popular ways to achieve the rapid response to frequent changes in software artefacts \cite{Senapathi2018,66MeedeniyaIGI2020}. DevOps practice is an emerging software development methodology that bridges the gap between the development and operations teams. DevOps eases the project team management with better communication, understandability, integration and relationships by bridging the gap between the development and operational teams. 
The integration of machine learning (ML) practices that support data engineering, with the DevOps based software development, has resulted in Machine Learning Operations (MLOps). It incorporates ML models for solution development and maintenance with continuous integration to provide efficient and reliable service. Different roles such as data scientists, DevOps engineers, and IT professionals are involved in this process. As an extension of DevOps with ML, the concept of MLOps increases collaboration and supports efficient software development by following a Continuous Integration Continuous Delivery (CICD) pipeline \cite{30PalihawadanaMERCon2017,46RubasingheICACCI2018}. It aims to produce software effectively and efficiently and supports software evolution. 

Accordingly, adhering to multiple project management tools has become a bottleneck in MLOps as both the task allocation and communication among teams is equally important in their highly collaborative nature. Thus, the MLOps environments have had to maintain a large tool stack to facilitate them. The manual process of maintaining consistency during the development process is a tedious and time-consuming approach \cite{Ashmore2019}. Consequently, automated traceability maintenance among heterogeneous software artefacts following a proper Change Impact Analysis (CIA) and Change Propagation (CP) in an MLOps environment that is focused on frequent artefact changes is challenging. These existing limitations have motivated this research study.

The core research question addressed in this study is exploring tool support in the MLOPs process. Thus, the goal of this research is to present widely used MLOps platforms with a collaborative environment that facilitates iterative data exploration, real-time co-working capabilities for experiment tracking, synchronization, feature engineering, and model management, together with controlled model transitioning, deployment, and monitoring. We provide an analysis of functionalities available in these tools for the benefit of data scientists and software engineers, which is a hindrance in related studies, as a novel contribution. This study discusses the available MLOps platforms, for the benefit of researchers and developers in this area. Thus, the MLOps teams can select the most suitable platform to satisfy their requirements.

The paper is structured as follows. Section II presents background related to DevOps and ML life cycle since a combination of those two concepts builds up MLOps technological stack. Then, Existing platforms are critically analyzed and compared in Section III. Moreover, a comparison of such platforms and weaknesses are described in Section IV. Section V concludes our survey study with suggestions and possible future research directions for MLOps.

\section{Background}
\label{sec:headings}

\subsection{Overview of DevOps}

DevOps-based software development supports to speed up the delivery time and frequency of delivering while supporting quality, reliability, and security \cite{Leite2019}. Currently, it has been the trend of many organizations to automate their delivery while keeping it as a bridge that connects software development and software deployment combining development and operations teams. Thus, DevOps is a continuous process including continuous development, integration, continuous deployment, and monitoring \cite{55MeedeniyaIJACSA2019}.

DevOps consists of a stack of support tools to detect a change in a software artefact and manage the consistency among other artefacts that are affected by the change to ensure the efficiency and fast delivery of solutions \cite{56MeedeniyaIJACSA2019,93RubasingheJIPS2021}. Different Integrated Development Environments (IDE) are used during the software development process. There are supporting mechanisms such as Git for version controlling and a docker container for packaging with all the libraries and dependencies. Tools such as Jenkins supports the integration of codes that are committed by multiple users. Thus, different tools support the frequent and reliable software releases with CICD \cite{Manish2015}.

Generally, the software artefacts change due to various reasons such as change of client’s requirements, socio and technological aspects. The automation tool support to detect these artefact changes, estimate the impact of the change, maintain the consistency among artefacts are required. Those tools should support better visualization and communication functionalities, as the DevOps process is collaborative \cite{41RubasingheICIS2018,30PalihawadanaMERCon2017}. The availability of automation tools and technical competencies supports reducing the effort required during the software development process \cite{87MeedeniyaICICM2021}; hence avoiding unnecessary costs associated with it. 

\subsubsection{Overview of MLOPs}

\begin{figure}[h!]
\begin{center}
\includegraphics[width=9.0cm]{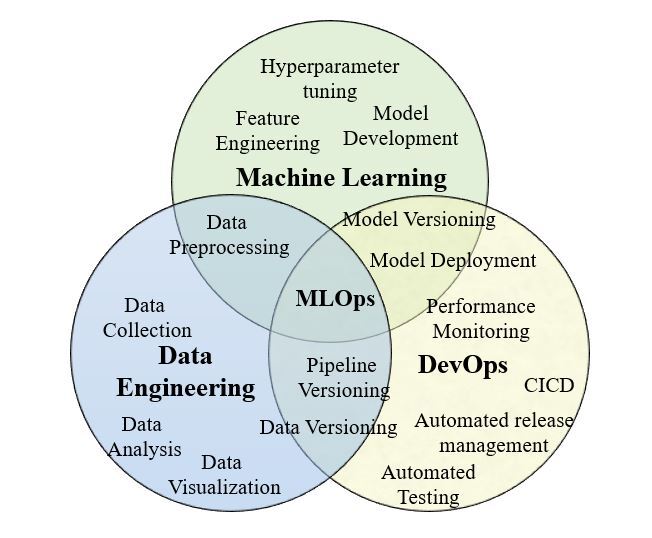}
\caption{MLOps combination}
\label{MLOps-combination}
\end{center}
\end{figure}

The MLOps practice brings ML models into the software solution production process. It bridges the ML applications with DevOps principles, where deployments and maintenance of ML models can be automated in the production environment as shown in Figure \ref{MLOps-combination}. MLOps systems should be capable of acting as collaborative, continuous, reproducible, tested, and monitored systems to achieve organizational MLOps goals. The development life cycle of MLOps consists of three major components as data, model, and code \footnote{Machine learning operations [online]. Website \href{https://ml-ops.org/}{https://ml-ops.org/} [accessed on 06 December 2021]}.

    \hfill \break

The following tasks of the MLOps framework require tool automation to maintain the life cycle  \footnote{MLOps Infrastructure Stack [online]. Website \href{https://ml-ops.org/content/state-of-mlops}{https://ml-ops.org/content/state-of-mlops} [accessed on 06 December 2021]}.
\begin{itemize}
    \item Data engineering tasks (collection, analysis, cleaning)
    \item Version Controlling for data, model, and code for model development
    \item CICD pipelines for process automation
    \item Automated model deployments and test automation
    \item Performance assessing, reproducibility, and monitoring in production to find deviations
\end{itemize}
Accordingly, MLOps provides efficient, scalable software development with reduced risk. Efficiency is achieved by fast model development, providing high-quality ML models, rapid deployment and production. The MLOps tool allows to oversee, control, manage and monitor many models with CICD supporting scalability. The high collaboration among teams reduces conflicts and accelerate the releases. Further, traceability and consistency management help to reduce the associated risks.

\subsection{Machine learning lifecycle}

The integration of ML with software development is accompanied by CICD. These training models are associated with different factors such as algorithms, hyperparameters that are tuned recursively. Once the model is deployed, it should be continuously monitored to avoid any performance degradation. Different roles, skill sets, and tools are utilized during the development life cycle. The ML Life cycle forms multiple stages such as model requirement, data collection, and preprocessing, design and development, evaluation and deployment, and monitoring. 

\begin{figure}[h!]
\begin{center}
\includegraphics[width=18.0cm]{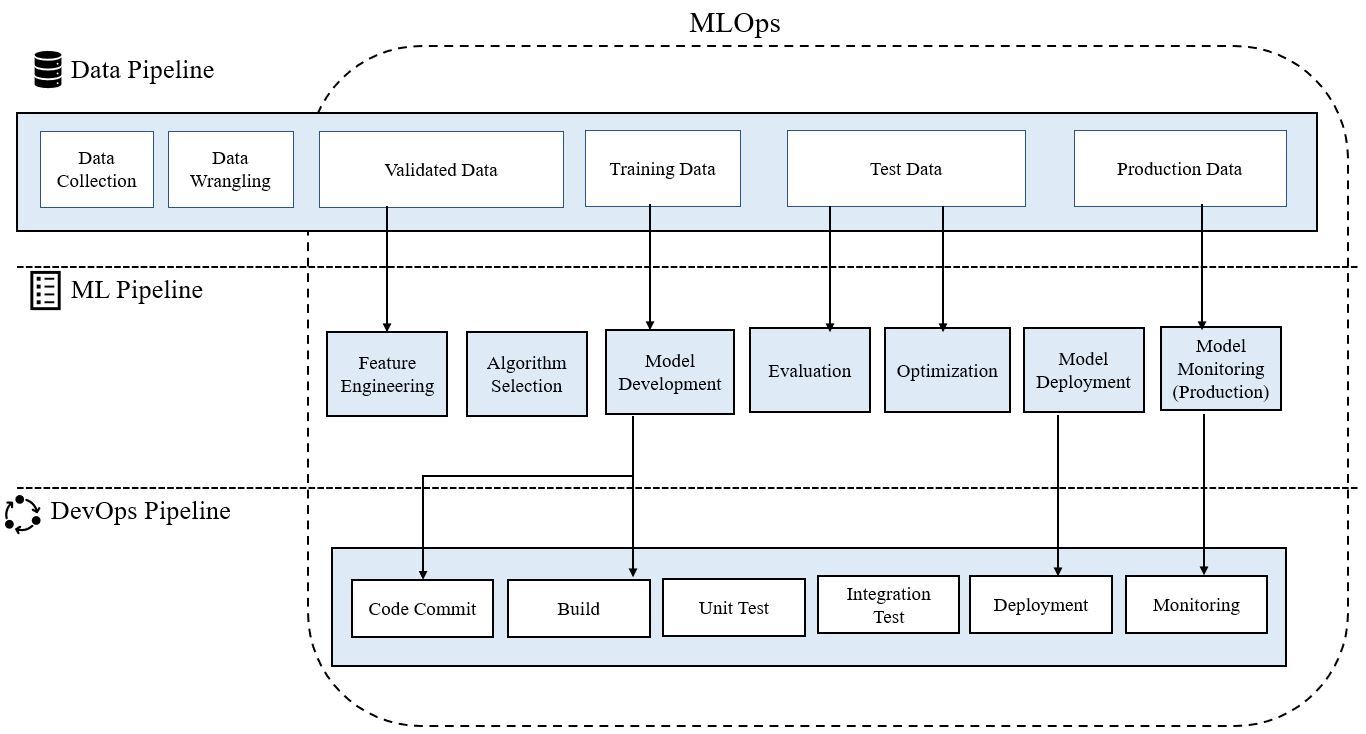}
\caption{High-level process view of MLOps}
\label{High-level process view of MLOps}
\end{center}
\end{figure}

Data is the factor, that determines the overall effectiveness of an ML model \cite{Paleyes2020}. Data can be open-source or private and collected using surveys or experiments. Due to the inaccuracy and redundancy of data, they should be cleaned and preprocessed before using for training \cite{Spjuth2021}. Then feature engineering techniques are applied to extract and identify vital informative features for the design of the ML models \cite{Amershi2019}. Hyper-parameter tuning and optimization processes are implemented before the training process. A repository is maintained to manage models and codebase. Once the code is committed to the repository, the code build stage is triggered with the DevOps practices. Unit testing and integration testing stages will be accompanied by the code build stage using a pipeline. Model testing and validation are also important to check the performance of the model. When the model complies with the expected accuracy level, it is deployed to the production environment. Continuous monitoring should be conducted similar to traditional software development. Figure \ref{High-level process view of MLOps} shows the layered interaction view of the data pipeline, ML pipeline, and DevOps pipeline associated with the MLOps practice \cite{mlopsIntro}. Thus, MLOps is defined as a procedure for automating the ML life cycle by reducing the human intervention in repetitive processes.

\begin{figure}[h!]
\begin{center}
\includegraphics[width=13.0cm]{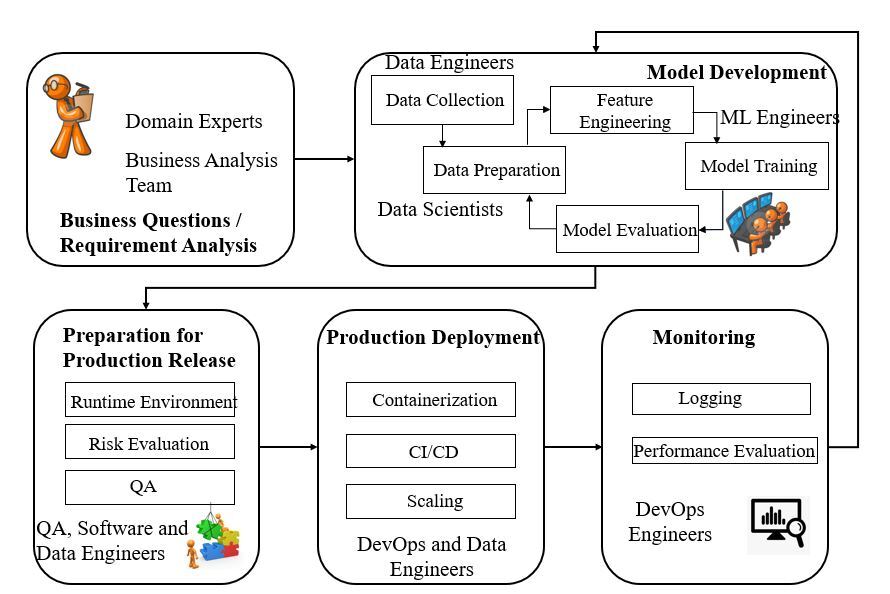}
\caption{MLOps phases and associated roles}
\label{MLOps phases and associated roless}
\end{center}
\end{figure}

Generally, the MLOps cycle commences with the business questions and requirement analysis done by the domain experts and business analysis team. Depending on the requirements, the designers decide on the type of models which is to be developed, the features that need to be considered, and the data collection mechanism and their availability \cite{mlopsIntro}. Accordingly, different roles such as data engineers, data scientists, and software engineers are occupied in each of these phases to accomplish the goals as shown in Figure \ref{MLOps phases and associated roless}. Before the production-level release, software engineers, data engineers and quality assurance engineers work on the factors that are inevitable in doing a release. Production development is carried out by DevOps and Data engineers using DevOps practices and tools considering scaling factors, security, and reliability measurements. Finally, Continuous monitoring and evaluation are performed with the use of DevOps techniques. 

\section{MLOps Tool Stack}
\label{sec:others}
\begin{figure}[h!]
\begin{center}
\includegraphics[width=15.0cm]{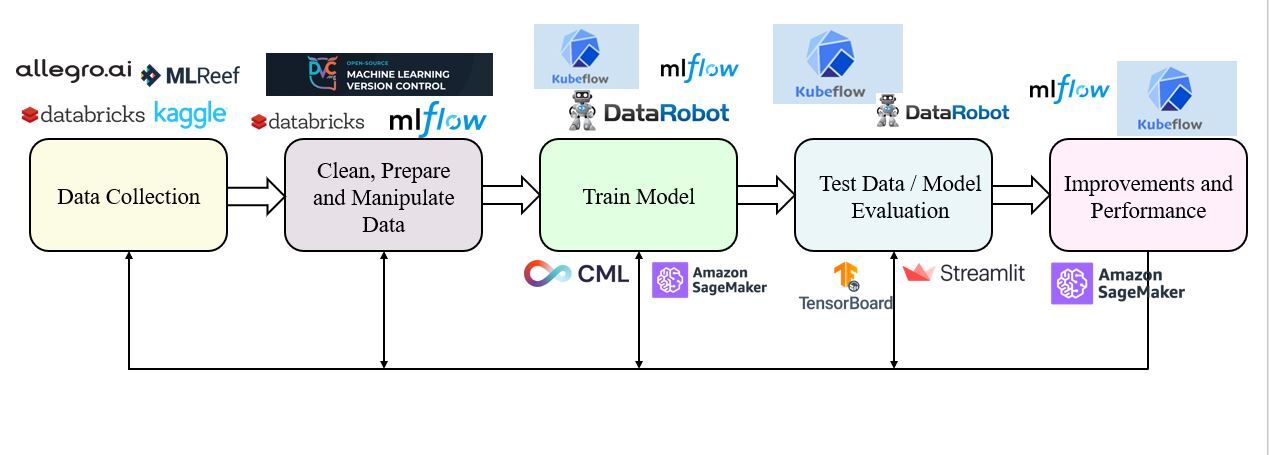}
\caption{MLOps tool stack }
\label{MLOps tool stack}
\end{center}
\end{figure}

MLOps tool stack supports managing the ML life cycle easier, reliable with fast delivery. As shown in Figure \ref{MLOps tool stack} many tools in the stack are utilized to accomplish one or more phases and this section discusses widely used tools in practice. 

\subsection{Kubeflow}
\begin{figure}[h!]
\begin{center}
\includegraphics[width=14.0cm]{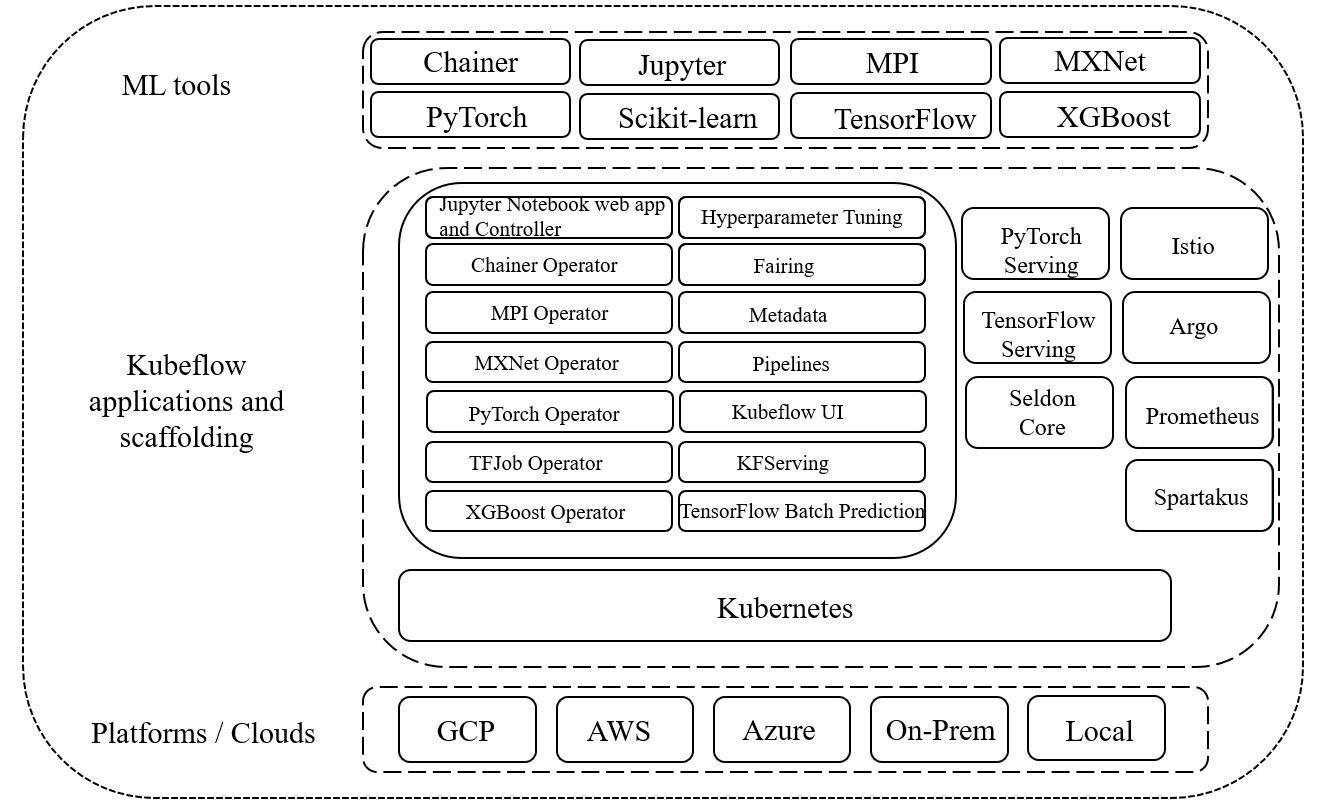}
\caption{Kubeflow model Architecture \protect\footnotemark}
\label{Kubeflow model Architecture}
\end{center}
\end{figure}

The Kubeflow project started at Google is committed to deploying ML models by allowing Kubernetes to manage simple, portable, and scalable deployments according to the requirements \footnote{Kubeflow [online]. Website \href{https://www.kubeflow.org/}{https://www.kubeflow.org/} [accessed on 06 December 2021]}. It is an open-source ML platform to organize the artefacts of the ML system on top of the Kubernetes system and supports to development, deployment, monitor throughout the life cycle of an ML application using automated pipelines. The conceptual diagram of Kubeflow is shown in Figure \ref{Kubeflow model Architecture}. Kubeflow facilitates a few ML frameworks and plugins for monitoring. It also comes with an interactive user interface (UI), notebook servers, Kubeflow pipelines, KFServing (model deployment and serving toolkit), training operators. Although currently, it does not have a dedicated tool for the CICD process, the Kubeflow pipelines can be used to construct reproducible work plans that automate the steps needed to build an ML workflow, which delivers consistency, saves iteration time, and helps in debugging, and compliance requirements \footnote{MLOps Infrastructure Stack [online]. Website \href{https://ml-ops.org/content/state-of-mlops}{https://ml-ops.org/content/state-of-mlops} [accessed on 06 December 2021]}.

\footnotetext{Kubeflow [online]. Website \href{https://www.kubeflow.org/}{https://www.kubeflow.org/} [accessed on 06 December 2021]}

\subsection{MLFlow}

MLFlow is an open-source, non-cloud platform for managing the end-to-end ML lifecycle \footnote{An open source platform for the machine learning lifecycle [online]. Website \href{https://mlflow.org/}{https://mlflow.org/} [accessed on 06 December 2021]} tackling the four primary functions: tracking, projects, models, and model registry in MLflow. MLflow Tracking allows users to track experiments to record and compare parameters and results by keeping logs and queries of all the inputs, versioning, and outputs \footnote{MLflow Tracking [online]. Website \href{https://mlflow.org/docs/latest/tracking.html}{https://mlflow.org/docs/latest/tracking.html} [accessed on 06 December 2021]}. The MLflow project can be used as a packaging tool for ML code which packages the code in a reusable and reproducible manner \cite{Zaharia2018}. MLflow models manage many ML libraries \footnote{MLflow Models [online]. Website \href{https://mlflow.org/docs/latest/models.html}{https://mlflow.org/docs/latest/models.html} [accessed on 06 December 2021]} and deploy to model serving and interference platforms. The MLflow model registry section helps the central mode store to jointly govern the full life cycle of an ML model including versioning, phase changes, annotations \footnote{MLflow Model Registry [online]. Website \href{https://mlflow.org/docs/latest/model-registry.html}{https://mlflow.org/docs/latest/model-registry.html} [accessed on 06 December 2021]}. Importantly, MLflow handles and executes any ML library and any programming language. Moreover, it allows to deploy and serve models as a web service with the usage of AWS SageMaker, Azure ML, Apache Spark \footnote{MLflow Models [online]. Website \href{https://mlflow.org/docs/latest/models.html}{https://mlflow.org/docs/latest/models.html} [accessed on 06 December 2021]}, achieving CICD goals via cloud service functionalities. Further, it supports statistical performance monitoring of deployed models \cite{Banerjee2020}. However, not having in-built notebooks and not maintaining notebook versioning to be used as IDE for the development are limitations in this tool. In addition, MLFlow does not maintain user management and does not offer full customizability like grouping experiments \cite{ruf2021}. 

\subsection{Iterative Enterprise}

The iterative enterprise consists of Data Version Control (DVC), Continuous Machine Learning (CML) and support, that manage and operate ML models, datasets, and experiments. Data versioning is a vital role in MLOps and it is challenging to handle when the dataset is large. DVC is an open-source platform-independent versioning system for ML applications and capable of creating ML models in a shareable, reproducible manner, while keeping versions for models, data, and pipelines. Additionally, it can generate small metafiles to support and keep track of large files, data sets, models, experiment data, metrics, and code to make maximum use of versioning \footnote{What is DVC? [online]. Website \href{https://mlops-guide.github.io/Versionamento/}{https://mlops-guide.github.io/Versionamento/} [accessed on 06 December 2021]} \footnote{Open-source version control system for machine learning projects [online]. Website \href{https://dvc.org/}{https://dvc.org/} [accessed on 06 December 2021]}. CML facilitates CICD for ML projects. It depends on GitLab or GitHub actions to manage ML experiments, keep track of modifications, auto-generate reports with metrics and plots in each Git pull request \footnote{Continuous Machine Learning (CML) is CI/CD for Machine Learning Projects [online]. Website \href{https://cml.dev/}{https://cml.dev/} [accessed on 06 December 2021]}. Additionally, DVC studio allows effective collaborative knowledge sharing among teams.

\subsection{DataRobot}
\begin{figure}[h!]
\begin{center}
\includegraphics[width=14.0cm]{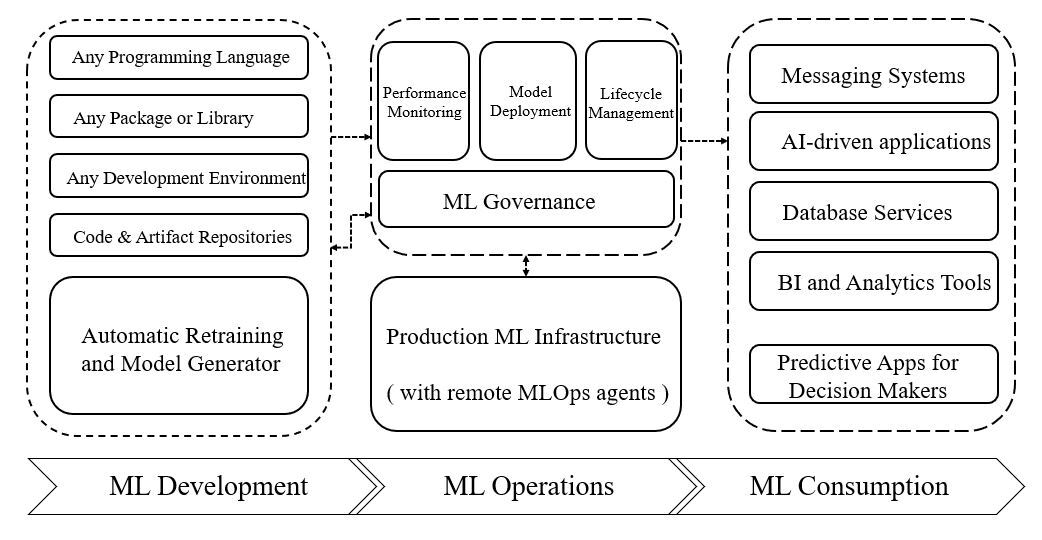}
\caption{DataRobot tool architecture \protect\footnotemark}
\label{DataRobot tool architecture}
\end{center}
\end{figure}

The DataRobot MLOps platform supplies a single place to deploy, monitor, manage models in productions regardless of how they were created, when and where they were deployed  \footnote{Datarobot Docs -  MLOps [online]. Website \href{https://docs.datarobot.com/en/docs/mlops/index.html}{https://docs.datarobot.com/en/docs/mlops/index.html} [accessed on 06 December 2021]}. It has a model registry to store and manage all production deployed models. As shown in Figure \ref{DataRobot tool architecture}, from ML development to consumption, DataRobot facilitates ML life cycle stages. It also supports many programming languages, libraries, development environments and maintains code repositories. However, individual users are required to purchase licenses for each instance to embedded usages.
\footnotetext{Datarobot AI Cloud Platform [online]. Website \href{https://www.datarobot.com/}{https://www.datarobot.com/} [accessed on 27 December 2021]}

\subsection{Allegro.ai (ClearML)}
\begin{figure}[h!]
\begin{center}
\includegraphics[width=9.0cm]{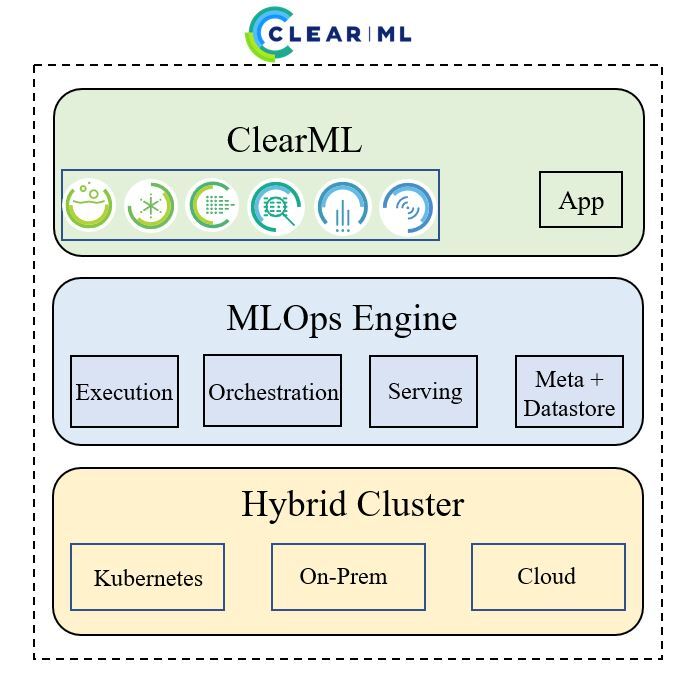}
\caption{ClearML stack architecture}
\label{ClearML stack architecture}
\end{center}
\end{figure}

Allegro.ai provides open-source MLOps tools to deliver products efficiently \footnote{allegroAI [online]. Website \href{https://www.allegro.ai}{https://www.allegro.ai} [accessed on 27 December 2021]}. ClearML is a product of Allegro.ai that enables a single place to experiment, orchestrate, deploy and build data store \footnote{ClearML [online]. Website \href{https://clear.ml/}{https://clear.ml/} [accessed on 27 December 2021]}. The main stages of ClearML are named as experiment, orchestrate, DataOps, hyper-datasets, deploy, and remote. Figure \ref{ClearML stack architecture} shows the model architecture and supports customizability. In addition, ClearML supports a set of modules. For instance, the ClearML python package integrates the codebase with the framework. ClearML Server consists of controlling features for MLOps while storing experiments, models, and workflow data. ClearML agent provides orchestration, reproducibility, scalability functionalities. ClearML session module provides remote instances of Jupyter Notebooks and VSCode.

\subsection{MLReef}
\begin{figure}[h!]
\begin{center}
\includegraphics[width=12.0cm]{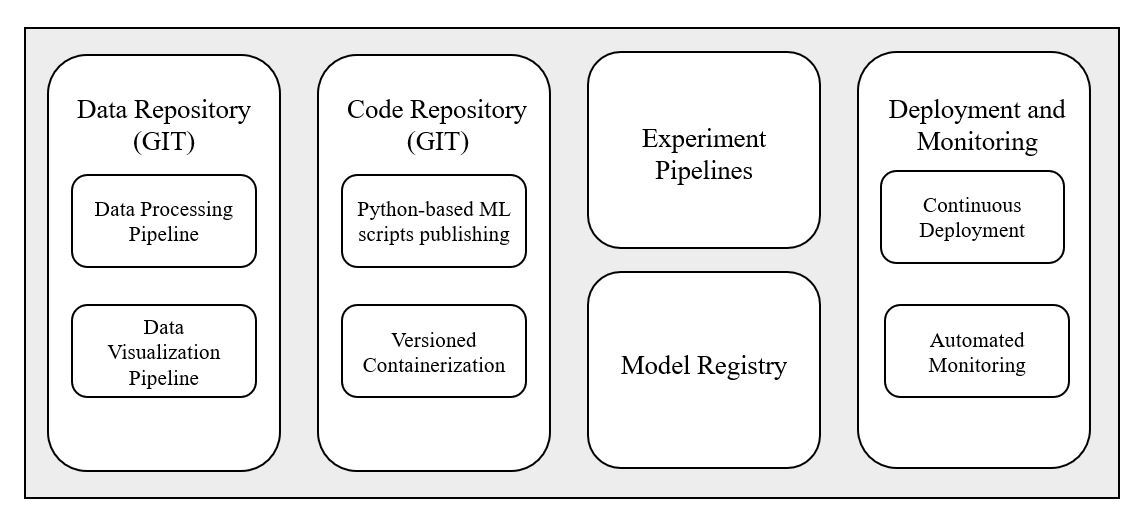}
\caption{MLReef architecture\protect\footnotemark}
% \footnote{MLReef [online]. Website \href{https://about.mlreef.com/}{https://about.mlreef.com/} [accessed on 27 December 2021]}
\label{MLReef architecture}
\end{center}
\end{figure}

MLReef is an open-source git-based MLOps platform, that offers a single location to manage the ML life cycle. In order to achieve reproducible, efficient and collaborative ML development, this platform manages work in repositories. Due to the capabilities of super-fast, collaboration, sharing, reproducibility, free CPU/GPU availability, and ownership, MLReef is better among MLOps platforms \footnote{MLReef [online]. Website \href{https://about.mlreef.com/}{https://about.mlreef.com/} [accessed on 27 December 2021]}. Figure \ref{MLReef architecture} shows the MLReef architecture that supports CICD. 

\footnotetext{MLReef [online]. Website \href{https://about.mlreef.com/}{https://about.mlreef.com/} [accessed on 27 December 2021]}
\subsection{Streamlit}
Streamlit is a python library that facilitates web app creation in less time. It has an easy to use interface without backend configurations \footnote{Streamlit [online]. Website \href{https://censius.ai/mlops-tools/streamlit}{https://censius.ai/mlops-tools/streamlit} [accessed on 27 December 2021]}. Streamlit supports code iteratively and views results while ongoing development. Users can deploy their web apps instantly using the in-built web server and visualize the behaviour using Streamlit cloud technology. Since it is a python library that gives a better impression for data visualization, it can be generally used for dashboard development-related tasks.

\subsection{MLOps with cloud service providers}
MLOps lacks mature solutions and hence it uses a set of tools to automate the process and requires frequent human interaction rather than DevOps. Generally, the cloud service providers offer ML platforms such as AI Platform by Google Cloud, AzureML studio by Microsoft Azure, Amazon SageMaker by Amazon Web Service (AWS) to increase the productivity of ML solutions. They also offer options to easily get used to ML for users who do not have proper knowledge of AI. The Pay-As-You-Go cost model for cloud services also encourages users to get associated with ML platforms provided by cloud service providers.
\textbf{Microsoft Azure} offers a set of components to support MLOps as follows \footnote{Machine learning operations (MLOps) [online]. Website \href{https://azure.microsoft.com/en-us/services/machine-learning/mlops}{https://azure.microsoft.com/en-us/services/machine-learning/mlops} [accessed on 06 December 2021]}.
\begin{itemize}
    \item Azure ML: supplies the ability to build, train, and validate myriads of models on daily basis regardless of the skill set, consists of built-in Notebooks
    \item Azure Pipelines: automates ML pipelines to build and test codes
    \item Azure Monitor: tracks and analyzes metrics to improve performance
    \item Azure Kubernetes Service
\end{itemize}

The setting up of the MLOps environment on Google cloud service provides a set of functionalities as follows \footnote{Setting up an MLOps environment on Google Cloud [online]. Website \href{https://cloud.google.com/architecture/setting-up-an-mlops-environment}{https://cloud.google.com/architecture/setting-up-an-mlops-environment} [accessed on 06 December 2021]}.
\begin{itemize}
    \item Dataflow: data-management service which extracts, transform data, and then evaluate models using data
    \item AI platform notebook supplies a development area to develop models (e.g.: Managed Jupyter notebook instances)
    \item Cloud Build: build, test, and deploy applications
    \item TFX: capable of deploying ML pipelines
    \item Kubeflow pipelines: automating ML deployments on top of Google Kubernetes Engine
\end{itemize}
MLOps on AWS can be achieved via \textbf{Amazon SageMaker}, a fully functional framework that can manage the ML lifecycle by automating MLOps practices. It supports to development, training, testing, deployment and monitor ML applications efficiently and productively.

\section{Discussion}
\subsection{Comparison of MLOps tools}
Research has been done to explore new horizons on developing sophisticated systems for MLOps. Although several tool support is available to manage the artefact traceability DevOps practice \cite{67RubasingheIGI2020}, there are no major tools that address the traceability in the MLOps life cycle. Several studies have presented automation tools to maintain the artefact consistency during the DevOps-based software development \cite{34RubasingheICCIP2017,46RubasingheICACCI2018}. The concept of these tools can be used to manage the traceability of the MLOps as well. 
Few studies have discussed the inevitability of MLOps due to the barriers and hectic manual processes that need to be improved frequently \cite{Makinen2021}. Most of the existing studies have addressed the development of MLOps technology frameworks \cite{Fursin2020}. Some of the commercially available platforms like MLflow, kubeflow are also capable of providing those functionalities up to some extent with automated processes. Also, these tool support will ease the development process and estimate the needed efforts. Table \ref{table1} states a comparison of functionalities addressed by the existing MLOps platform. The features data versioning (DV), hyperparameter tuning (HT), model and experiment versioning (MEV), pipeline versioning (PV), CICD availability, model deployment (MD) and performance monitoring (PM) were considered for the comparison of the MLOps platforms. This can be referred to when selecting a suitable platform for the solution development environment. 

\begin{table}[!ht]
\caption{Feature comparison of existing platforms}
\begin{center}
\begin{tabular}{|l|l|l|l|l|l|l|l|}
\hline
&
DV & HT & MEV & PV &  CICD & MD & PM \\

\hline
AWS SageMaker & 
\checkmark & 
\checkmark & 
\checkmark & 
\checkmark & 
\checkmark & 
\checkmark & 
\checkmark  \\
\hline
MLFlow & 
\checkmark& 
\checkmark& 
\checkmark& 
\checkmark& 
\checkmark& 
\checkmark& 
 \\
\hline
Kubeflow &
& 
\checkmark& 
\checkmark& 
& 
\checkmark& 
\checkmark& 
\checkmark  \\
\hline
DataRobot &
& 
\checkmark& 
\checkmark& 
& 
& 
\checkmark& 
\checkmark \\
\hline
Iterative Enterprise &
\checkmark& 
& 
\checkmark& 
& 
\checkmark& 
\checkmark& 
\checkmark  \\
\hline
ClearML  &
\checkmark& 
& 
\checkmark& 
\checkmark& 
\checkmark& 
\checkmark& 
\checkmark  \\
\hline
MLReef  &
\checkmark& 
\checkmark& 
\checkmark& 
\checkmark& 
\checkmark& 
\checkmark& 
\checkmark  \\
\hline
Streamlit  &
\checkmark& 
& 
\checkmark& 
& 
& 
\checkmark& 
\checkmark  \\
\hline
\end{tabular}
\label{table1}
\end{center}
\end{table}

Moreover, software development environments use different programming languages, libraries, and frameworks. Thus, an MLOps platform should be capable to supply services in a platform-independent manner. In that case, when choosing a better MLOps platform, supporting languages, frameworks and libraries should be taken into consideration. Table \ref{table2} summarizes the languages supported by different MLOps platforms.
Accordingly, it can be seen that MLflow and AWS SageMaker perform better than others, but they also have weaknesses that need to be addressed as described under each of the frameworks previously. Although cloud service providers have similar platforms, they are costly and are not addressing the ML problem itself through a single dashboard. In addition to that, some of the platforms do not offer free licenses to use as embedded systems.

\begin{table}[h!]
\caption{Language-support comparison of existing platforms}
\begin{center}
\begin{tabular}{|l|l|l|l|l|l|l|l|l|}
\hline
&
PyTorch &
Jupyter  & 
Java &
TensorFlow & 
Scikit- & 
Keras &
R &
Python \\
&
&
Notebook &
&
&
learn
&
&
\\

\hline
AWS SageMaker & 
\checkmark& 
\checkmark& 
\checkmark& 
\checkmark& 
\checkmark& 
\checkmark& 
\checkmark&
\checkmark \\
\hline
MLFlow & 
\checkmark& 
& 
\checkmark& 
\checkmark& 
\checkmark& 
\checkmark& 
\checkmark& 
\checkmark \\
\hline
DataRobot &
\checkmark& 
& 
\checkmark& 
\checkmark& 
\checkmark& 
\checkmark& 
\checkmark&
\checkmark \\
\hline
Kubeflow &
\checkmark& 
\checkmark& 
& 
\checkmark& 
\checkmark& 
& 
& 
\checkmark   \\
\hline
Iterative Enterprise &
& 
& 
& 
\checkmark& 
\checkmark& 
& 
& 
\checkmark  \\
\hline
ClearML  &
\checkmark& 
\checkmark& 
& 
\checkmark& 
\checkmark& 
\checkmark& 
& 
\checkmark  \\
\hline
MLReef  &
\checkmark& 
& 
& 
\checkmark& 
\checkmark& 
\checkmark& 
&
\checkmark  \\
\hline
Streamlit  &
\checkmark& 
& 
&
\checkmark& 
& 
\checkmark& 
& 
\checkmark  \\
\hline
\end{tabular}
\label{table2}
\end{center}
\end{table}

\subsection{Current challenges and future research directions}
The accuracy of the predictions made by ML applications depends on many factors such as data type, training algorithm, hyperparameters, learning rate and optimizers. Some applications such as precipitation models need the latest real-time data and are retrained frequently to produce more accurate and precise predictions. Thus, the training models should be retrained without human intervention using reproducible pipelines. It is challenging to automate these decisions making processes using MLOps. In addition, the MLOps platform should be capable of creating workflows, models and allowing them to be reused and reproduced easily to expand the area of the experiments to reach the expected performance. ML pipelines can be provided to develop, deploy and redevelop using the previously designed models for faster and reliable delivery of the solutions. Dataset registries and model registries can be managed and maintained so that they can be reused and available for modifications to varying data sets in the future. 
Continuous training and evaluation techniques and strategies might be beneficial to have in such kind of platform. The platform should be capable of migrating accurate and confidentially packed models into production easily and allowing auto-scaling according to the needs of CPU, GPU metrics. CICD can be applied to straightforwardly accomplish such requirements as in DevOps. In addition, the design and development of supporting tools to automate the MLOps process can be extended by incorporating natural language processing (NLP) as well \cite{22ArunthavanathanMERCon2016}. The health of the models should be live monitored and precautions should be taken to reduce the impact on the production application. Moreover, these MLOps platforms should be user friendly, reliable, and efficient to use in practice.

\section{Conclusion}
This survey study explored the importance of Machine Learning Operations (MLOps), and the functionalities and limitations of available platforms which in turn directs researchers to expand their horizons for the development of sophisticated similar platforms. We emphasized the need for usable and efficient tool support to maintain the consistency between artefacts in software development practices that involve machine learning models, continuous integration and DevOps. This survey study compared commercially available MLOps platforms which can be used to fulfil the needs of the ML life cycle. Although several MLOps platforms are available in practice, most of them have limitations to accomplish ML life-cycle phases delivering an automated framework. The analysis of the available platforms opens a new research direction to develop a fully automated user interface based MLOps dashboard that can be used by domain experts and developers. 
%Bibliography
%\bibliographystyle{unsrt}  
%\bibliographystyle{plainnat}
\bibliographystyle{IEEEtranN}

\bibliography{MLOps}

\end{document}